\documentclass{vgtc}                          

\ifpdf
  \pdfoutput=1\relax                   
  \usepackage{graphicx}                
  \DeclareGraphicsExtensions{.pdf,.png,.jpg,.jpeg} 
\else
  \usepackage{graphicx}                
  \DeclareGraphicsExtensions{.eps}     
\fi%

\graphicspath{{figures/}{pictures/}{images/}{./}} 

\usepackage{microtype}                 
\PassOptionsToPackage{warn}{textcomp}  
\usepackage{textcomp}                  
\usepackage{mathptmx}                  
\usepackage{times}                     
\usepackage{cite}                      
\usepackage{tabu}                      
\usepackage{booktabs}                  

\onlineid{0}

\vgtccategory{Research}

\vgtcinsertpkg

\title{A Mixed-Initiative Visual Analytics Approach for Qualitative Causal Modeling}

\author{Fahd Husain \thanks{e-mail: [fhusain, pproulx, dchang, rgomez, hvasquez]@uncharted.software} %
\and Pascale Proulx %
\and Meng-Wei Chang %
\and Rosa Romero-Gómez %
\and Holland Vasquez} %

\affiliation{\scriptsize Uncharted Software Inc.}

\teaser{
  \centering
  \includegraphics[width=0.9\linewidth]{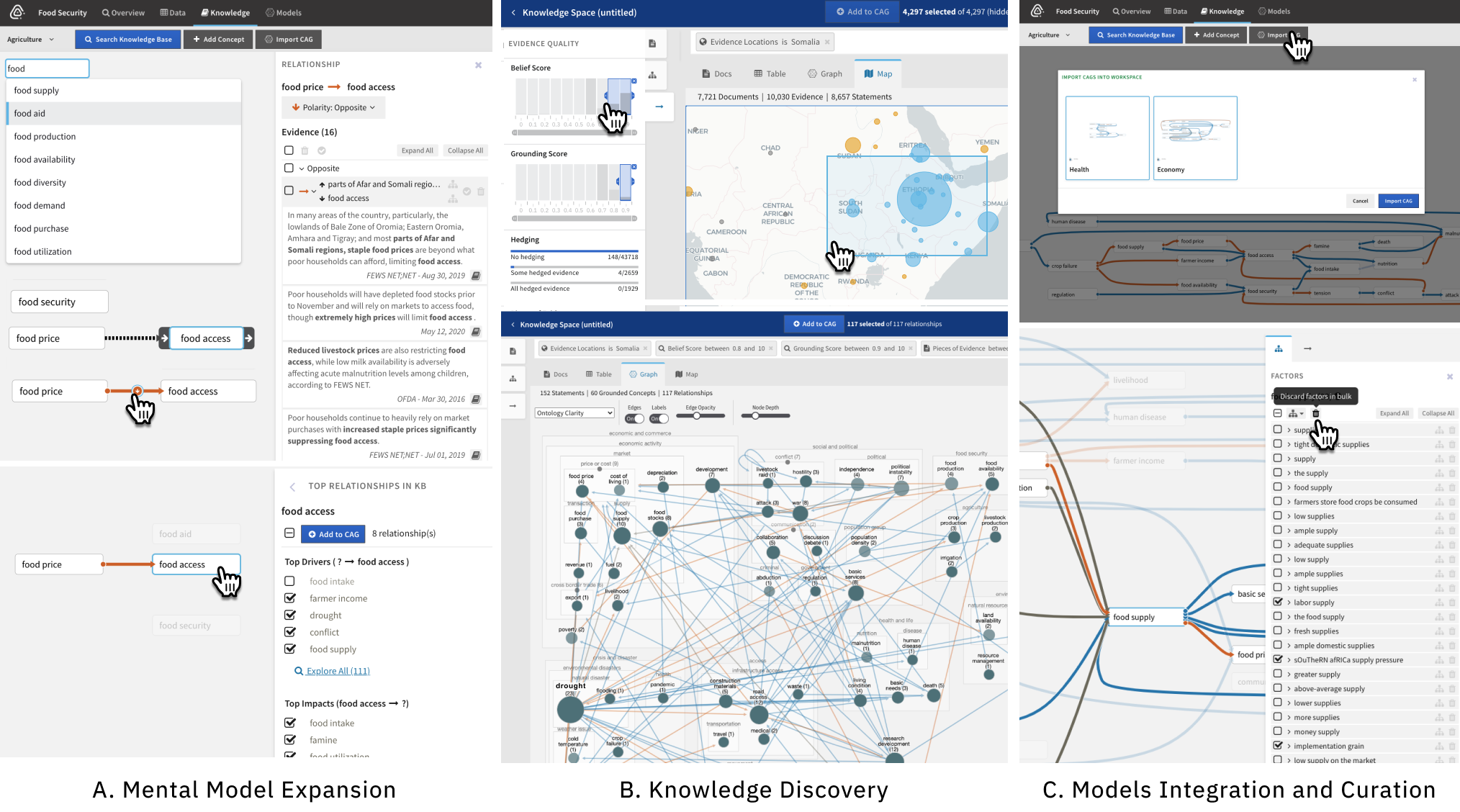}
  \caption{Our mixed-initiative visual analytics approach, a subset of the Causemos platform, accelerates the ability of analysts to assemble qualitative models of complex socio-natural systems by leveraging causal relationships automatically extracted from domain literature. \textbf{A. Mental Model Expansion} shows how the analyst captures prior knowledge about the key relationships relevant to food security in East Africa as a causal analysis graph (CAG). 
  \textbf{B. Knowledge Discovery} shows a faceted explorer of the underlying knowledge base. The top map displays East Africa as a user-selected region of focus. The bottom graph displays a set of filtered causal statements using a hierarchical graph layout that organizes graph nodes by domains, such as  economy, policy and health. \textbf{C. Models Integration and Curation} shows how two different models related to economy and health can be merged together with an agricultural model to form a more comprehensive model. Ambiguous polarities and near-duplicate nodes are curated in bulk by the analyst using system recommendations.}
  \label{fig:teaser}
}

\abstract{Modeling complex systems is a time-consuming, difficult and fragmented task, often requiring the analyst to work with disparate data, a variety of models, and expert knowledge across a diverse set of domains. Applying a user-centered design process, we developed a mixed-initiative visual analytics approach, a subset of the Causemos platform, that allows analysts to rapidly assemble qualitative causal models of complex socio-natural systems. Our approach facilitates the construction, exploration, and curation of qualitative models bringing together data across disparate domains. Referencing a recent user evaluation, we demonstrate our approach's ability to interactively enrich user mental models and accelerate qualitative model building.%
} 

\CCScatlist{
  \CCScatTwelve{Human-centered computing}{Visu\-al\-iza\-tion}{Visu\-al\-iza\-tion techniques}{Treemaps};
  \CCScatTwelve{Human-centered computing}{Visu\-al\-iza\-tion}{Visualization design and evaluation methods}{}
}

\begin{document}


\maketitle
\section{Introduction}
The study of complex systems remains a critical research area impacting many domains in the social and natural sciences. For instance, in the era of climate change, it is crucial to model interconnected events such as famine, food security, and political instability. However, the dynamics of such phenomena are typically poorly understood, and solutions are either lacking consensus or simply unavailable \cite{Rittel:1973}. 
Against such complexity, systems methodology employs interdisciplinary thinking to articulate these phenomena in terms of systems and subsystems, boundaries and environments, relationships and feedback loops \cite{Bertalanffy:1969, Checkland:1976}. Modeling causality also provides insight into system structure and dynamics, as well as internal forces responsible for change \cite{pearl:2009}. 
Understanding the interactions between natural processes and human systems poses significant challenges, and currently requires working with siloed models, disparate data, and expert knowledge across disciplinary boundaries\cite{Gil:2018}. Many models are manually built \textit{ad hoc} for new situations, requiring many months of effort, and resulting in long delays for critical events that need immediate interventions \cite{Sharp:2019}. For example, Foresight’s obesity report\cite{Butland:2007} sought to shed light on the complex causal structures around obesity in order to inform public health decision making efforts. 
It took one hundred multi-disciplinary experts six months to gather data, assemble models, and complete intervention policy studies. Such efforts highlight that manually constructing qualitative causal models leaves organizations hard-pressed to provide quick and effective intervention guidance in complex real-world scenarios. 


The World Modelers (WM) program\cite{WorldModelers:2019} aims to address these challenges in order to accelerate the capability of analysts to model complex socio-natural systems. The principal human-machine interface (HMI) for the WM program is \textit{Causemos}, a visual analytics platform designed to enable the efficient assembly and execution of causal models by analysts. The primary research contribution of this paper is a mixed-initiative visual analytics approach for the interactive exploration, curation, and assembly of qualitative causal models. As part of the Causemos platform, this approach leverages multi-dimensional faceting, nested and flow graph layouts, and system suggestions in various contexts.
We demonstrated the efficacy of our approach in a user evaluation in which analysts worked to model food security in East Africa. Analysts were able to browse and filter knowledge from thousands of documents in minutes instead of weeks and expand the scope of their mental model. The approach allowed transparency, which fostered analysts' trust in the platform's results. 

\section{Related Work}
In modeling complex systems, many tools employ a qualitative approach  to causal modeling, during which users create a graph that visually describes the factors represented as nodes and influences of those factors for a particular problem. Factors are attributes of an entity or their environment that influence the question of interest \cite{Butland:2007}. Relationships between factors are commonly displayed as a set of links \cite{Landesberger:2011, pearl:2009}.

Qualitative causal modeling tools such as argument mapping (e.g. \cite{wright2017argument}), mind mapping (e.g.\cite{chen2019mini}), and knowledge graphs (e.g.\cite{Hirsch2009, milanlouei2020systematic} are primarily focused on providing a digital workspace to allow users to sketch their mental models. However, most of these tools only rely on the user's knowledge and experience and do not provide support for expanding the user's mental model, validating the user's prior knowledge with evidence, or integrating work from multiple analysts to facilitate discussions  \cite{chen2019mini, kapler2021causeworks}. Additionally, current systems are limited with respect to the speed with which users can build their model. In many instances, users manually construct qualitative causal models, which can be a time-consuming process \cite{Butland:2007, Hatfield:2015, milanlouei2020systematic}. 
There are a number of tools that automatically extract causal relationships from data sets \cite{chen_data_2011, krueger_vespa_2017, lu_visual_2018, xie2020causalflow, kapler2021causeworks}, that allow analysts to visualize causal relationships and interact with the visualizations. However to our knowledge, most causal modeling tools are limited to providing the analyst with a bottom-up approach to modeling \cite{krueger_vespa_2017, kapler2021causeworks}. 

Our approach has many parallels with the systems above, keeping a graph as the basis for visualizing the  qualitative causal model. Causemos allows analysts to quickly capture their mental model by automatically identifying supporting evidence from the literature and augmenting the user's mental model with in-context system suggestions. What sets our work apart is the ability to both visually explore the automatically-mined literature and leverage a mixed-initiative knowledge curation approach that mitigates automatic knowledge extraction noise. 
Ultimately, by augmenting these capabilities, we allow analysts to rapidly produce more comprehensive evidence-backed qualitative causal models. 

\section{Design Process and Objectives}
We followed a user-centered design approach that engaged with analysts, modeling experts, and government agencies from both North America and East Africa. Early in the design process, we relied on structured and semi-structured interviews and observations of analysts and modeling experts performing modeling tasks related to food security. The interviews provided insight into how modeling efforts support local government decision-making around natural resources (agriculture), health (disease, infection), socio-economics (poverty, instability), and climate change (rainfall, drought). We observed that in modeling complex socio-natural systems, analysts often need to take on a multidisciplinary approach, and in doing so, frequently have to tackle issues for which they lack domain expertise. As such, analysts often need to conduct literature reviews to identify relevant factors and relationships. To do that, analysts may rely on information gathered from agency and non-governmental organization reports, empirical studies, surveys, field observations, or published datasets. 
When faced with short time frames to conduct modeling efforts, analysts will often limit the scope of their analysis to a small set of key factors. Additionally, analysts will work together as a team and divide tasks among team members. However, a key challenge is integrating the results from multiple analysts into a cohesive model that can be used to inform policy and decision-making efforts. 


In light of the insights gleaned from our interviews and observations, we distilled four high-level design objectives (DO) that were used to inform the design and implementation of our qualitative model assembly approach:

\begin{itemize}
	\setlength\itemsep{1em}
    \item  \textbf{DO1. Allow rapid assembly of qualitative causal models}. 
    Analysts should be able to assemble a qualitative causal model of a complex system in the form of a graph in a matter of minutes, which may ultimately accelerate policy and decision-making. 
      
    \item \textbf{DO2. Support the expansion of the analyst's mental model to allow for multi-domain analysis}. The analyst should be able to learn about factors related to their problems that may be outside of their specific domain expertise. 
    
  \item \textbf{DO3. Complement machine-assistance with trust to assemble models}. 
  Provenance information needs to be easily accessible and quality issues easily detectable and effectively addressed. As such the analyst should be able to synthesize a comprehensive analysis that can be presented with confidence to policy and decision makers.
  \item \textbf{DO4. Support for the integration of models.} In order to work collaboratively while leveraging different modeling perspectives, analysts should be able to integrate individually-developed qualitative causal models into a single, coherent, and more comprehensive model.
    
\end{itemize}

\section{Visualization Design}

\subsection{Causal Analysis Graph (CAG)}

\begin{figure}[tb]
\centering 
\includegraphics[width=8cm,height=4cm,keepaspectratio]{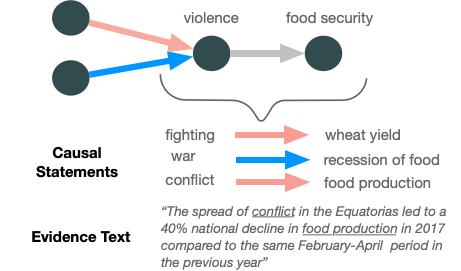}
\caption{Aggregation schema of causal statements. Extracted causal statements (with underlying evidence) are rolled up into the graph edge ``violence \textrightarrow food security''.
An example of the evidence-supporting the statement between conflict and production is shown.}
\label{fig:causal_statements_aggregation}
\end{figure}

A qualitative causal model in Causemos is represented in the form of a directed acyclic graph (DAG) composed of a set of causal statements with nodes representing concepts and edges representing causal relationships (see Fig.\ref{fig:teaser}.A, bottom left). These causal statements are aggregated at the concept level (see Fig.\ref{fig:causal_statements_aggregation}). Causal assertions are extracted by collaborating machine-reading and knowledge assembly systems\cite{Sharp:2019} resulting in a set of normalized, qualitative causal statements organized by a curated ontology\cite{wmontology:2020}. The polarity of a causal statement is either ``same'' (i.e. a change in the subject causes the object to change in the same direction) or ``opposite''. We use a color-blind safe two-color scale to encode the polarity of an edge, blue for ``same'' and red for ``opposite''. However, since many causal statements may be aggregated under an edge, sometimes an edge can have ambiguous polarity, which is encoded as gray (see Fig.\ref{fig:causal_statements_aggregation}). This is a cue for the analyst to review the evidence since the ambiguity is likely to stem from a quality issue (\textbf{DO3}). We also surface and aggregate the belief score of the statements, and encode it with the edge's saturation as another indication of the quality of the relationship. Belief score is defined as the overall support for a statement given the joint probability of correctness implied by the evidence\cite{Sharp:2019}.

Since directionality is critical for understanding causality, we use a familiar left-to-right flow layout algorithm\cite{Sugiyama:1981} that has been extended  more recently \cite{elkjs:2020} by assigning x and y-coordinates separately in multiple stages. The layout is also adaptive based on the size of the graph, with slightly reduced spacing constraints for larger graphs to offset the number of ``lanes`` created by the layout algorithm. The graph will start off centered and fitted to the canvas. The analysts can then perform \textit{zooming} and \textit{panning} operations to view parts of the model in greater detail.



\subsection{Sketching Mental Models}
Causemos allows analysts to directly capture their mental model while being supported by existing assembled knowledge in the form of causal statements (\textbf{DO1, DO2}). The analyst starts this process either by double-clicking on the graph background or by clicking the ``Add Concept'' button (see Fig.\ref{fig:teaser}.A, top left). This operation displays a search box in context along with a suggestion list showing the closest matching concepts ranked by the amount causal statements in the system about these concepts. Hovering over a node reveals dragging controls used to draw an edge to another node. To avoid edge crossings, we use the A* algorithm \cite{astar} to do edge routing on the fly.

\subsubsection{Understanding Evidence and Curation}
Causemos automatically surfaces causal statements that match the source and target concepts and adds them to the edge. By selecting the edge, the textual evidence is visible in the side panel (see Fig.\ref{fig:teaser}.A, top right) with the option to open the source document to see the evidence in context (\textbf{DO3}). Here an analyst can gauge the quality of the machine extractions, should there be any errors or disagreements. Causemos provides bulk editing actions to correct concept assignments, change polarity, or discard statements altogether (\textbf{DO3}). In cases where no statements can be found, Causemos can crawl causal statements as knowledge graphs and come up with indirect suggestions (\textbf{DO1, DO2}). We achieve this by pre-computing clusters \cite{hdbscan} in the embedding space represented by the causal statements \cite{Glove}. For example, it may be the case that there is no direct causal relationship between ``disease'' and ``farming'', but there is an indirect relationship from ``disease'' to ``livestock'', and another from ``livestock'' to ``farming''. This feature serves two diametric purposes: it can reveal a gap in the underlying knowledge when analysts know a relationship should exist but it does not, and it also offers alternative ideas that analysts may not have considered before because they tend to operate in a siloed domain. Causemos leaves it up to the analysts to decide if they want to use the indirect paths between concepts, or if they want to create an edge with no backing statements, in which case the edge stroke switches to a dotted black line to denote no evidence and unknown polarity.

\subsubsection{Model Extension Based on Suggestions}
In addition to drawing relationships, the CAG can also be extended relationship-by-relationship in a piece-wise manner. Selecting a node will open the side panel to show the current incoming and outgoing relationships. Clicking the ``Suggestions'' button in the side panel presents the top five relationships coming into and going out of the node ranked by the number of supporting causal statements in the knowledge base (see Fig.\ref{fig:teaser}.A, bottom right). The analysts can quickly augment their model by selecting one or more of these relationship suggestions (\textbf{DO1, DO2}). 

However, it is sometimes the case that the analysts will disagree with the composition of a particular edge given its aggregated nature. To handle this case, Causemos provides a toggle option in the side panel (see Fig.\ref{fig:teaser}.A, bottom left) to selectively pick whether the edge aggregated polarity is ``same'' or ``opposite'' (\textbf{DO3}). Likewise, an edge with no backing statements can be configured in the same manner to allow full expressivity. Removal of unwanted nodes and edges is simply accomplished by selecting the corresponding graph elements and pressing the ``delete'' key. 

\subsection{Leveraging the Knowledge Base}
An alternate method to build all or part of a qualitative causal model is to contextualize the problem at hand as a faceted search query, which can be conducted in the Knowledge Explorer view (see Fig.\ref{fig:teaser}.B). 
This view allows analysts to perform faceted search across multiple levels (\textbf{DO1, DO2, DO3}): 1) at the source document level (e.g. return only causal statements from a specific paper published between 2010 and 2015); 2) at the relationship level (e.g. return only statements with opposite polarities that are backed by at least three pieces of evidence); and 3) at the factor and statement geo-temporal context level (e.g. return only statements where the subject or object matches one of the concepts: ``flood'', ``education'', or ``conflict'', and whose geographical context is within Eastern Africa).

The combination of these three levels of search abstraction allows the analysts to express the context of their problem. The search results are shown in different representations to allow better understanding of the data they are about to use (\textbf{DO3}). A tabular view provides textual information of the underlying causal statements. An interactive map view (see Fig.\ref{fig:teaser}.B, top) provides filtering capabilities as well as the context in which events have occurred. Lastly, a nested graph visualization (see Fig.\ref{fig:teaser}.B, bottom) shows both the relationship structure and relative placement of concepts in the ontology. We employ an adapted hierarchical layout algorithm\cite{Dogrusoz:2009} that has been successfully used to visualize biological pathways with nesting and compartmental constraints. To avoid edge cluttering, we hide edges when there are more than 2,000 relationships in the search results. Here the analysts can merge the search results back into the CAG, or they can fine-tune this search by selecting specific edges from the nested graph.

\subsection{Integrating Existing Models}
Causemos offers the capability to import existing models into the analyst's workspace. The primary motivation for this is to accommodate the siloed working style of analysts while offering the chance to collaborate and leverage each others' domain knowledge (\textbf{DO1, DO4}). Several small teams can be working on different problems or different dimensions of a problem that can be combined together to form a more cohesive and representative picture. To achieve this, analysts can click the ``Import'' button, which will show a list of the other existing models (see Fig.\ref{fig:teaser}.C, top). Analysts can then select one or more models to import into their own. Ambiguous polarities and near-duplicate nodes can be curated in bulk by the analyst using system recommendations (see Fig.\ref{fig:teaser}.C, bottom). 

\section{Usage Scenario}
Sara, a fictional analyst, is trying to identify the factors and relationships that are most likely to cause food shortages in East Africa.
She starts by entering key factors that she knows are relevant to the problem. As she types ``food'' in the search box, Causemos provides suggestions of food-related concepts, ranked by amount of knowledge in the system (see Fig.\ref{fig:teaser}.A, top left). From the list, she picks food security, food price, food access, and food aid. Then, she draws a relationship between food price and food access, which immediately turns the edge stroke from a dashed line to a solid red edge. She opens the side panel to reveal extracted statements from multiple documents that present evidence of increases in food price leading to reduce food access (see Fig.\ref{fig:teaser}.A, top right). Sara also looks at the suggested relationships for food access on the side panel. Selecting from this list is a faster way to flesh out her mental model, leveraging recognition rather than recall (see Fig.\ref{fig:teaser}.A, bottom right). 
After spending a few more minutes to add other concepts and relationships, her CAG contains about thirty factors, spanning mostly socio-political domains, which reflects her expertise. She decides to open the Knowledge Explorer view and filter the existing knowledge to identify other relevant factors using the nested graph (see Fig.\ref{fig:teaser}.B, bottom). Given the large size of the ``drought'' node, which reveals many causal statements relevant to the East African region (see Fig.\ref{fig:teaser}.B, bottom), she quickly realizes she needs to add it to her CAG. To include health and economic factors, Sara decides to import the CAGs created by her colleague (see Fig.\ref{fig:teaser}.C, top right).
By reviewing the underlying evidence of these models, Sara notices several causal statements about labor supply are incorrectly mapped to a food supply node. She uses the bulk-correction action to rectify this mapping (see Fig.\ref{fig:teaser}. C, bottom). In the end, Sara produces a comprehensive model and reaches her goal of identifying major factors leading to food system shortages.

\section{Evaluation}
Twenty East African technical analysts participated remotely (i.e., through a video call) in a user evaluation with the goal of assessing the efficacy of our approach in the context of the Causemos platform. This evaluation lasted five days with sessions consisting of training (three, two-hour sessions) and hands-on use (approximately 2-3 hours each day). Participants had between 10 and 30 years of experience in analyzing food security-related issues, and had diverse skills and areas of expertise including data analytics, engineering, agriculture, economics, water resources, public health, and climate change. The participants received no compensation for taking part in the study. 

Participants were presented with fictional scenarios in which their objective was to model and forecast food security in East Africa using key indicators centered on food availability (e.g., cereal crop production), access (e.g., household income), utilization (e.g., malnutrition rates), and stability (e.g., likelihood and impact of climate and health shocks). Analysts also had to identify optimal interventions to mitigate these scenarios with the goal of informing policymakers. At the end of each day, participants were asked to respond to a series of questions related to the efficacy of Causemos and to provide open-ended feedback on the system. Responses to open-ended questions were analyzed in a thematic analysis. An affinity diagram \cite{harboe2015real} was used to extract themes in the data. 
In what follows, we describe the five main themes identified in user feedback. 

\textbf{Contributions and System Utility}: Ten analysts commented on the contributions and utility of Causemos. Rather than taking weeks to read documents, extract relationship networks and isolate hypotheses, analysts were able to interactively browse and filter knowledge from 40,000 documents and build qualitative causal models within 30 to 60 minutes. Analysts noted that Causemos enabled broader scope analysis than was usually untenable in their typical six-week time frame, such as considering the relationship between fuel costs and foreign currency reserves, and their link with food security. 

\textbf{Trust in the System}: Five analysts reported several features that bolstered their trust in Causemos. For instance, all relationships are backed by evidence and the system is transparent in that it provides the analyst with references to the identified supporting evidence. However, four analysts also reported that there were barriers to trusting the system. For example, analysts noted that there were some instances where evidence was thin or non-existent and expressed uncertainty in how the system extracts information from machine reading.   

\textbf{Requested Data and System Features}: Eleven analysts noted that there was a need for additional data to complete their analysis. For instance, some analysts desired additional academic articles, peer-reviewed qualitative research, and specific data related to disasters and interventions. Importantly, analysts requested additional information related to system functionality. For example, analysts requested definitions for terminologies used in the knowledge search, details on how the various technologies are integrated in the system, and a user manual. 

\textbf{Challenges and frustrations}:
Nine analysts mentioned experiencing challenges or frustrations when performing their analysis. For example, analysts noted that not all expected concepts were available in the knowledge base, some relationships lacked evidence or were ambiguous, and curating evidence was a time consuming process.

\textbf{A Need for Practice and Training}: 
Five analysts indicated that the training provided was insufficient and that they desired more hands-on practice with Causemos. In line with this, several analysts noted that training manuals and additional details related to the various technologies integrated in the system would be helpful in supporting novice users.

\section{Conclusion and Future Work}
Modeling complex socio-natural systems in a comprehensive and timely manner remains a challenge for organizations attempting to provide guidance for effective interventions on critical situations. Our mixed-initiative visual analytics approach addresses this challenge by supporting analysts in the assembly, exploration, analysis, curation, and integration of qualitative causal models of such complex systems.  Positive feedback from analysts underscores the efficacy of our approach. 

In light of the evaluation, we are extending Causemos to allow analysts to include better in-context suggestions and help, and to allow analysts to bring their own document data. We also plan to address visual scalability issues for the nested graph visualization by applying collapse-expand operations on demand as well as visual salience measures for relevant nodes and edges. Finally, we are exploring ways to extend the modeling workflow towards building quantitative models that can be simulated and support \textit{what-if} questions to explore optimal interventions.

\acknowledgments{
We wish to thank DARPA’s World Modelers Program (BAO: HR001117S0017) for supporting this research. The views, opinions, and findings contained in this paper are those of the authors and should not be construed as an official Department of Defense position, policy, or decision. This work has been approved for Public Release, Distribution Unlimited. The authors wish to thank all World Modelers collaborators for their support and encouragement.}

\bibliographystyle{abbrv-doi}

\bibliography{template}
\end{document}